# Effects of MAC Parameters on the Performance of IEEE 802.11 DCF in NS-3

Md. Abubakar Siddik[1,*], Jakia Akter Nitu[1], Natasha Islam[1], Most. Anju Ara Hasi[1], Jannatun Ferdous[1], Md. Mizanur Rahman[2] and Md. Nahid Sultan[3]

[1]Department of Electronics and Communication Engineering, Hajee Mohammad Danesh Science and Technology University, Dinajpur, Bangladesh
[2]Department of Electrical and Electronics Engineering, Hajee Mohammad Danesh Science and Technology University, Dinajpur, Bangladesh
[3]Department of Computer Science and Engineering, Hajee Mohammad Danesh Science and Technology University, Dinajpur, Bangladesh

## *Abstract*

*This paper presents the design procedure of the NS-3 script for WLAN that is organized according to the hierarchical manner of TCP/IP model. We configure all layers by using NS-3 model objects and set and modify the values used by objects to investigate the effects of MAC parameters (access mechanism, $CW_{min}$, $CW_{max}$ and retry limit) on the performance metrics viz. packet delivery ratio, packet lost ratio, aggregated throughput, and average delay. The simulation results show that RTS/CTS access mechanism outperforms basic access mechanism in saturated state, whereas the MAC parameters have no significant impact on network performance in non-saturated state. A higher value of $CW_{min}$ improves the aggregated throughput in expense of average delay. The tradeoff relationships among the performance metrics are also observed in results for the optimal values of MAC parameters. Our design procedure represents a good guideline for new NS-3 users to design and modify script and results greatly benefit the network design and management.*

## 1. Introduction

WLANs have become increasingly popular in the past two decades as a primary and fast-growing wireless connectivity among wireless devices because of their facile and distributed nature connectivity, low network construction and maintenance cost, and simple implementation [1, 2, 3]. The IEEE 802.11 standard [4] defines the MAC and PHY layer specifications for connectivity with fixed, portable, and moving stations within a wireless local area network(WLAN).The IEEE 802.11 supports two contention-based medium access protocols, named as DCF and EDCA, use carrier sense multiple access with collision avoidance (CSMA/CA) as random access mechanism. The DCF has been widely used due to less complexity, simple operation and distributed nature. There are two access mechanisms: two-way handshake, basic access mechanism and four-way handshake, request-to-send/clear-to-send (RTS/CTS) mechanism used by DCF. The operation of CSMA/CA mechanism is controlled by several MAC parameters viz. minimum contention window ($CW_{min}$), maximum contention window ($CW_{max}$), DCF inter frame space (DIFS), and retry limit [5].

According to DCF, a station (STA) node with a packet to transmit continuously sense the channel activity. If the channel is sensed idle for a predefined time interval called DIFS, the STA node transmits the packet immediately. Otherwise, if the channel is sensed busy, the STA node initiates the backoff procedure. During a backoff procedure, the STA node will start a backoff counter with an initial value randomly selected from $[0, CW_{min} − 1]$ and continuously sense the channel. If the channel is idle for DIFS time, the backoff counter will resume. Then if the channel is sensed idle for CSMA time slot, the backoff counter will be decremented by one. When the backoff counter value reaches zero, the STA node transmits the packet immediately. If the STA node receives ACK successfully, then the backoff procedure for a new packet is initiated. If the packet transmission is not successful, the STA node again initiates the backoff procedure with a new backoff counter value randomly selected from $[0, W_j − 1]$, where, $W_j = 2^j \times CW_{min}$ and $j$ is the number of retry. If the value of $W_j$ reaches $CW_{max}$, the packet will be dropped by STA node and STA node initiates a new backoff procedure for a new packet.

In this paper, we develop a NS-3 script for a WLAN in NS-3 [6] which consists of an access point (AP) node and a number of STA nodes forming a one-hop star topology. We assume that each node of the network uses IEEE 802.11a standard specifications as MAC and PHY layer, TCP/IP protocol stack as network and transport layer and on off application and packet sink application as application layer. Moreover, we set and modify the attributes of different layer objects to investigate the effects of MAC parameters (access mechanism, $CW_{min}$, $CW_{max}$, and retry limit) on performance metrics viz. packet delivery ratio (PDR), packet lost ratio (PLR), aggregated throughput, and average delay. We also show the impact of number of STA nodes on performance metrics. Most of the previous works focus on the investigation of the effects of MAC and PHY parameters on performance but, to the best of our knowledge, none of the research work has described the design procedure of NS-3 script in accordance with the instructions of TCP/IP model to their manuscript.

The rest of the paper is organized as follows: Section 2 summarizes the related works. In Section 3, configuration of WLAN based on TCP/IP network model in NS-3 is described in details. Section 4 derives the performance metrics based on flow monitor attributes of NS-3. The effects of MAC parameters on performance metrics and most important findings of this work are presented in Section 5. Finally, Section 6 concludes the paper and gives the future work outlines.

## 2. RELATED WORKS

There are several literatures that investigate the effects of different MAC and PHY parameters on the performance metrics through mathematical model and/or simulation. In [7], G. Bianchi proposed an analytical model based on Markov chain to analyze the performance of the IEEE 802.11 DCF. The author shows the effects of access mechanism, $CW_{min}$, retry limit, and number of stations on throughput. Y. Yin et al. [8] conducted a performance evaluation study of IEEE 802.11 DCF via both analytical model with Markov model and simulation with NS-3. Throughput is measured for various network scenarios viz. distinct conditions, varying system parameters, different access modes and network topologies. N. Shahinet al. [9] proposed a cognitive backoff mechanism that adaptively determines the CW and evaluated the performance of proposed backoff algorithm with existing mechanisms (BEB, EIED, and ECA) using NS-3. They investigated the effects of access mechanism, initial contention window, retry limit, and number of nodes on throughput and delay. In [10], the authors evaluated the performance of IEEE 802.11 DCF via analytical model using Markov chain and simulation in OPNET. They present the effect of access mechanism, minimum CW, retry limit (finite and infinite), and number of nodes on aggregated throughput and access delay. A.Bozkurt [11] investigated the impact of retry limit, contention window, and velocity on throughput for IEEE 802.11 based

VANET using Markov chain based analytical model with M/G/1/k queue. S.Manzoor et al. [1] analyzed the performance of IEEE802.11 DCF through analytical model, Linux based test bed and verified by two simulation tools (OMNeT++, NS-3). They represent the effect of access mechanisms, backoff parameters (cutoff phase, window size), and number of nodes on optimal throughput. Y. Lee et al. [12] designed an analytical model based on Markov chain to evaluate the performance of IEEE 802.11 DCF under non-saturated condition and developed a C++ simulator modeling to verify the proposed analytical model. The results present the effects of packet arrival rates and number of nodes on normalized throughput. In [13], the authors explored the effects of access mechanism, number of nodes, minimum CW on throughput, and average delay of communication based train control (CBTC) through simulation. T. Kim et al. [14] proposed a new analytical model to accommodate the effect of the hidden terminal and validated this analytical model through simulation using NS-2. The results show the higher retry limit offers high aggregated throughput in presence of hidden terminal. Alkadeki et al. [15] proposed dynamic control backoff time algorithm (DCBTA) for CSMA/CA of IEEE 802.11 that outperforms BEB and ELBA when the network is in non-saturated state. An analytical model is developed for the per-node throughput analysis of IEEE 802.11 WLAN with hidden terminals by extending Bianchi's model in [16]. The validation of proposed analytical model is performed through simulation. The authors investigate the effects of access mechanisms, and minimum CW on per-node throughput and unfairness in per-node throughput due to hidden nodes. Jun Peng [17] developed Markov chain based analytical model to estimate the saturation throughput of an IEEE 802.11 network in the basic access mode and validated the proposed analytical model through simulation using NS-2. In [18], a simple analytical model is proposed to evaluate the throughput, delay and fairness in single-hop IEEE 802.11 networks and performance evaluation is carried out under three different Mac parameters viz. $CW_{min}$, $CW_{max}$, and retry limit. S. Lim [19] proposed an analytical model to set the optimum maximum CW to reduce collision and investigate the effects of maximum CW and retry limit on throughput.

## 3. SYSTEM CONFIGURATION IN NS-3

In this section, we outlet the configuration of WLAN based on TCP/IP network model in NS-3 and the measurement process of performance metrics using C++ language. Moreover, we present the hierarchical design procedure of WLAN in NS-3 and also highlight how the attributes or parameters of different NS-3 models or classes can modify to design a new configuration of a network. The overall communication architecture based on TCP/IP network model in NS-3 is given in Figure 2.

The NS-3 is a well-organized and maintained, flexible and simple architecture, easy and accessible documentation, fully open source, and widely used in academic and research as a network simulation tool and it has high accuracy and speedy execution capability to run NS-3 scripts. NS-3 provides different types of helper API, container API and core API to design a complete communication system. The container API performs a number of identical actions to groups of objects and the core API performs a particular task for a NS-3 model. The helper API is used to write and read NS-3 script easier. We first include the necessary namespaces and header files at top of the NS-3 script. The rest of the part of the NS-3 script is described as follows:

### 3.1. Physical Topology Configuration

In this work, we consider a WLAN consists of an AP node and a number of identical STA nodes in a star topology to investigate the effects of MAC and PHY parameters on the performance matrices. In order to create an AP node and a number of STA nodes, we take two objects of *NodeContainer* class, named as *apNode*, *staNode,* and use *create ()* function which takes the number of AP nodes and number of STA nodes as a parameter. To assign positions and to set mobility pattern to the AP and STA nodes, we take two objects of *MobilityHelper* class, named as *apMobility*, *staMobility,* and use *SetPositionAllocator ()* and *SetMobilityModel ()* functions. NS-3 provides different position allocators and mobility models to describe the network topology properly. Each position allocators and mobility models can set position and mobility pattern to the nodes according to a set of attributes. Finally, the defined properties of position allocator and mobility model are assigned to AP and STA nodes by using *Install ()* function which takes *apNode* and *staNode* objects as a parameters, respectively. In this work, we consider *ConstantPositionMobilityModel* class as a mobility model.

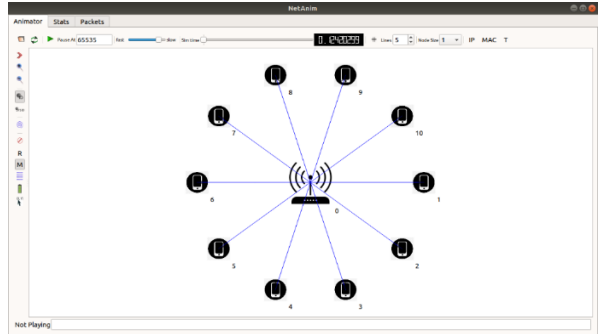

Figure 1: PHY topology of WLAN generated in NS-3

### 3.2. Network Device Configuration

The *NetDevice* class of NS-3 provides objects as network interface card of actual computer for Ethernet, WiFi, Bluetooth, etc. In this paper, we consider *WifiNetDevice* class to design IEEE 802.11-based WLAN. The *WifiHelper* class is used to create *WifiNetDevice* objects for each node which is the interface between network layers to data link layer. The common properties of *WifiNetDevice* like remote station manager and PHY standard for both AP and STA nodes are assigned by using *SetRemoteStationManager ()* and *SetStandard ()* functions through *WifiHelper* class object named wifi. To configure differentiable properties of *WifiNetDevice* for AP and STA nodes, we create two objects of *NetDeviceContainer* class, named as *apNetDevice* and *staNetDevice*. Finally, *WifiNetDevice* for AP and STA nodes is configured by using *wifi* object and *Install ()* function which takes *WifiPhyHelper* object, *WifiMacHelper* object and *NodeContainer* object as a parameters.

In this work, to show the effects of access mechanism, we modify the *RtsCtsThreshold* attribute value using *SetRemoteStationManager ()* function. We also set the retry limit by editing the *WifiRemoteStationManager* class and change the default value of *MaxSsrc* and *MaxSlrc* attributes to investigate the effects of retry limit.

### 3.2.1. PHY Layer Configuration

In order to configure *WifiPhy* model as the PHY layer, we take an object of Yans*WifiPhyHelper* class, named as *wifiPhy*, because both type of nodes (AP and STA) use same PHY layer. Moreover, we also assign error model to the PHY layer object by using*SetErrorRateModel ()* function. Then we take an object of *YansWifiChannelHelper* class, named as *wifiChannel*, to configure the channel for both AP and STA nodes. The *AddPropagationLoss ()* functions is used to set propagation loss model to the channel object and finally, this channel is assigned to PHY layer through *SetChannel ()* function.

### 3.2.2. MAC Layer Configuration

To configure MAC layer model for the AP and STA nodes, we take into account two objects of *WifiMacHelper* class, named as *apWifiMac*, *staWifiMac*, because AP and STA nodes need different configurations to communicate with each other. The NS-3 offers three types of MAC model like *ApWifiMac*, *StaWifiMac* and *AdhocWifiMac*. To set *ApWifiMac* model as the MAC layer of AP node, we use *SetType ()* function and assign different MAC parameters values as attributes to change the configuration as we needed. Similarly, configuration of MAC layer of STA nodes is performed by using *StaWifiMac* class *and SetType ()* function.

In this work, to show the impact of $CW_{min}$, and $CW_{max}$ on performance metrics, we assign different CW value by using *Set ()* function of *MatchContainer* class under *Config* namespace. It is remember that this *Set ()* function works after default configuration installation.

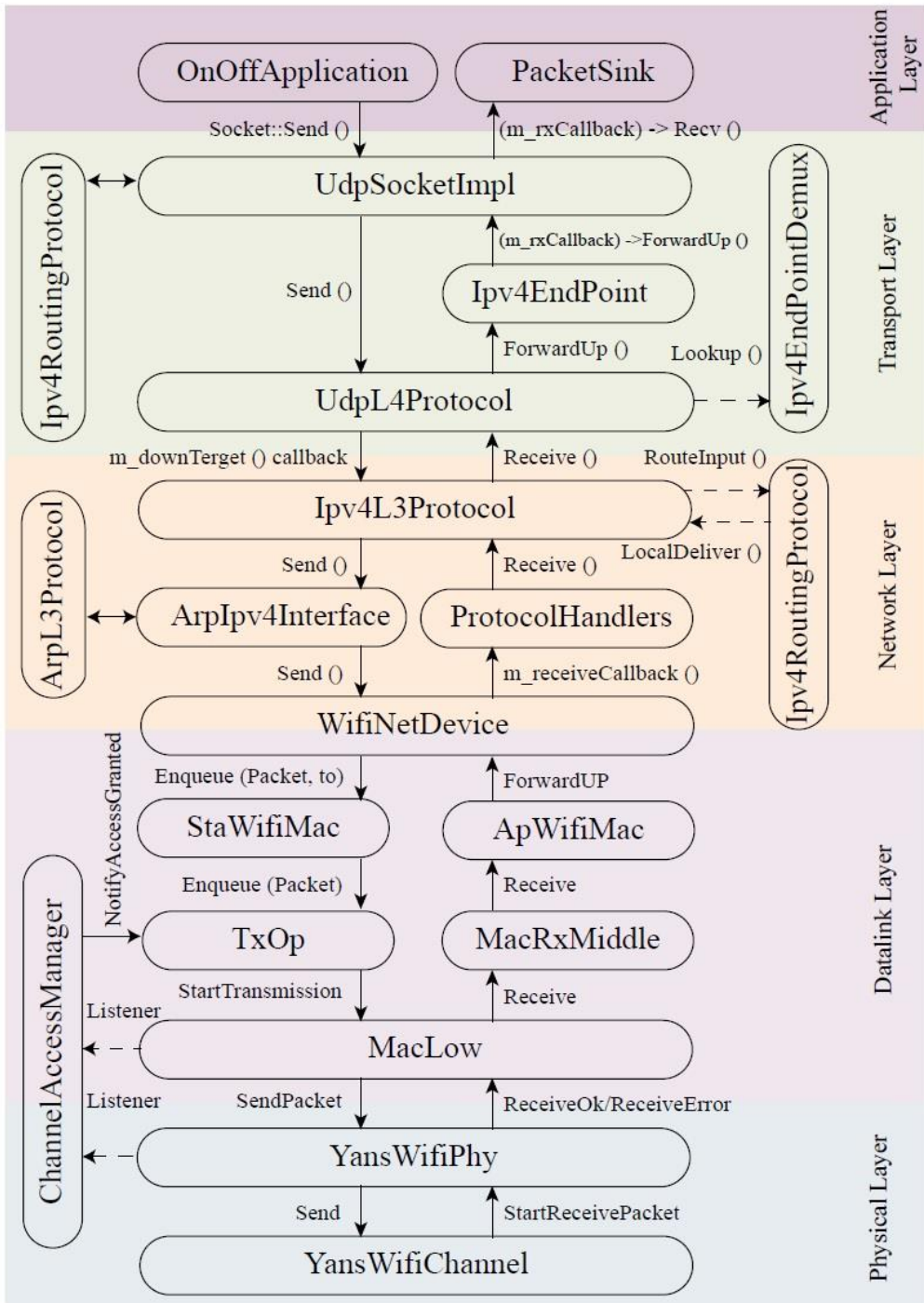

Figure 2: Communication architecture in NS-3

## 3.3. Network and Transport Layer Configuration

To configure network and transport layer of AP and STA nodes, we consider an object of *InternetStackHelper* class, named as internetStack, because both AP and STA nodes use same network and transport layer protocols. We set all default configurations of network and transport layer to *apNode*, *staNode* objects by using *Install ()* function. *InternetStackHelper* class aggregates IP/TCP/UDP functionality viz. *ArpL3Protocol*, *Ipv4L3Protocol*, *Ipv6L3Protocol*, *Icmpv4L4Protocol*, *Icmpv6L4Protocol*, *UdpL4Protocol*, *TrafficControlLayer*, *PacketSocketFactory*, *Ipv4* routing, *Ipv6* routing, by default, to each node. Finally, we finish the network layer configuration process after the end of the IP address set and assign procedure. In order to assign IP address to each network device of AP and STA nodes (apNetDevice and *staNetDevice*), we create an object of *Ipv4AddressHelper* class, named as *ipAddr*. We use *SetBase ()* and *Assign ()* functions to set IP address for AP and STA nodes and to assign this IP address through two objects of *Ipv4InterfaceContainer* class, named as *apNodeInterface* and *staNodeInterface*.

### 3.4. Application Layer Configuration

To configure application layer, we need application container where the protocols of application layer are installed. We take two objects of *ApplicationContainer* class, named as *apApplication* and *staApplication*. NS-3 offers different types of applications by defining different application class. In this work, we use *OnOffApplication* class as the application of STA nodes and *PacketSink* class as the application of AP node. Socket address (IP address of network layer protocol and port number of transport layer protocol) and transport layer protocol of the node are needed to configure an application. We take two objects of *InetSocketAddress* class, named as *apAddress* and *staAddress*, which contain the socket address of AP and STA nodes. Then, we consider an object of *PacketSinkHelper* class, named as *apSink*, and after that, *SetAttribute ()* function is used to define the attributes of the packet sink application. Finally, *Install ()* function is used to install the *PacketSink* application into the AP node. Similarly, we take an object of *OnOffHelper* class, named as *staOnOff*, and *SetAttribute ()* and *Install ()* functions are used to configure the *OnOffApplication* to each STA nodes. Finally, we initialize the start and stop time of the applications of AP and STA nodes by using *Start ()* and *Stop ()* functions of *ApplicationContainer* class.

### 3.5. Simulation Setup

The *Simulation* class of NS-3 executes the simulation events and control the virtual time. In this work, we use three static functions of *Simulation* class to execute our designed NS-3 script for *simulationTime* time. The *Run ()* function runs the simulation and will be continued until at least one of the three events occurs: (a) no events are present anymore, (b) user called *Stop ()* function, and (c) user called *Stop ()* function with a stop time and the expiration time of the next event to be processed is greater than or equal to the stop time. The *Destroy ()* function is typically called at the end of a simulation to avoid false-positive reports by a leak checker. After this method has been called, it is actually possible to restart a new simulation. The *Stop ()* function tells the *Simulator* class the calling event should be the last one executed. It takes the *simulationTime* as a function parameter.

### 3.6. Animation Setup

The NS-3 has two ways to provide animation, namely using the *PyViz* method or the *NetAnim* method. In this paper, we use *NetAnim* to display the topology of the network and animate the packet flow between nodes. The *NetAnim* also provides useful features such as tables to display meta-data of packets. To trace the file generated during simulation, we use an object of *AnimationInterface* class, named as *animation*. *AnimationInterface* class traces the statistics for each flow and exports in XML format.

### 3.7. Event Monitoring and Data Collection

The NS-3 supports many event monitoring model like ASCII trace, PCAP and flow monitor. In this work, we use flow monitor model to measure the performance of network protocols. The flow monitor is a flexible event monitoring model and it uses probes to track the packets at IP level exchanged by the nodes and the packets are divided according to the flow they belong to, where each flow is defined as the packets with same (protocol, source IP address, source port number, destination IP address, destination port number) tuple. Flow monitor collects the statistics for each flow and exports in XML format. To enable the flow monitor, we take an object of *FlowMonitorHelper* class, named as monitor and use *InstallAll ()* function. The data produced during simulation is traced by *FlowMonitor* class and flow monitor model stores this data in a map according to flow id using variables: *timeFirstTxPacket, timeLastTxPacket, timeFirstRxPacket, timeLastRxPacket, delaySum, jitterSum, txBytes, rxBytes, txPackets, rxPackets, timesForwarded, bytesDropped, packetsDropped*, etc. We manipulate this data and measure the performance metrics viz. PDR, PLR, aggregated throughput and average delay according to the definition defined in Section 4. After calculating performance metrics, we create a CSV file and write the value of performance metrics to the file using *ofstream* class of C++. Our designed NS-3 script is given in Appendix section in this paper.

## 4. PERFORMANCE METRICS

In this section, we outline how the principal performance metrics, viz. packet delivery ratio (PDR), packet lost ratio (PLR), aggregated throughput, and average delay, are determined from flow monitor attributes.

### 4.1. Packet Delivery Ration (PDR)

The packet delivery ratio is defined as the ratio of total number of packets received by AP for all flows and the total number of packets transmitted by STA nodes for all flows of the network. It is also measure from the total number of received bytes and total number of transmitted bytes. Therefore, if the number of STA nodes in the network is n, the PDR is expressed as:

$$\text{PDR} = \sum_{i=1}^{n} \frac{\text{Total received packets of flow i}}{\text{Total transmitted packets of flow i}} = \sum_{i=1}^{n} \frac{i-> second.rxPackets}{i-> second.txPackets} \quad (1)$$

### 4.2. Packet Lost Ration (PLR)

The packetlost ratio is defined as the ratio of total number of lost packets for all flows and the total number of packets transmitted by STA nodes for all flows of the network. Therefore, if the number of STA nodes in the network is n, the packet lost ratio is expressed as:

$$\text{PLR} = \sum_{i=1}^{n} \frac{\text{Total lost packets of flow i}}{\text{Total transmitted packets of flow i}} = \sum_{i=1}^{n} \frac{i-> second.lostPackets}{i-> second.txPackets} \quad (2)$$

### 4.3. Aggregated Throughput

The throughput of a STA node or a flow is defined as the number of bits of the STA node successfully received by AP node in unit time. Throughput is measured in bits per second (bps).

The aggregated throughput is the summation of indivudual STA node throughput. Therefore, if the number of STA nodes in the network is $n$, the aggregated throughput is expressed as:

$$\text{Aggregated throughput} = \sum_{i=1}^{n} \text{Throughput of flow i} = \sum_{i=1}^{n} \frac{\text{Received bytes of flow i}}{\text{Simulation time}}$$
$$= \sum_{i=1}^{n} \frac{i-> second.rxBytes \times 8}{simulationTime} \quad (3)$$

## 4.4. Average Delay

The average delay is defined as the ratio of the sum of all end-to-end delays for all received packets and total number of received packets. The average delay is measured in seconds (sec). Therefore, if the number of STA nodes in the network is $n$, the average delay is expressed as:

$$\text{Average delay} = \sum_{i=1}^{n} \frac{\text{Sum of packet delay of flow i}}{\text{Received packets of flow i}}$$
$$= \sum_{i=1}^{n} \frac{i-> second.delaySum.GetSeconds}{i-> second.rxPackets} \quad (4)$$

## 5. PERFORMANCE EVALUATION

In this section, we evaluate the effects of MAC parameters viz. access mechanism, $CW_{min}$, $CW_{max}$, and retry limit, on performance metrics of one-hope star topology structure in a WLAN under the IEEE 802.11a standard. The simulation experiments were conducted using NS-3 (version 3.30). In the simulation experiments, we considered four network scenarios (indexed by 1, 2, 3 and 4) that are used to investigate the effects of four MAC parameters separately. The simulation experiment consists of an AP node and a number of STA nodes that adopt all default configurations defined in NS-3 under IEEE 802.11a standard except few attributes which are given in Table 1.

Table1. Simulation attributes

| Configuration | Attributes | Values |
|---|---|---|
| General | Mobility model | *ConstantPositionMobilityModel* |
| | Remote station manager | *ConstantRateWifiManager* |
| | Data mode and control mode | *OfdmRate12Mbps* |
| | Wifi channel | *YansWifiChannel* |
| | Propagation loss model | *RangePropagationLossModel* |
| | Wifi PHY | *YansWifiPhy* |
| | Error rate model | *YansErrorRateModel* |
| | Wifi MAC for AP node | *ApWifiMac* |
| | Wifi MAC for STA nodes | *StaWifiMac* |
| | Number of AP nodes | 1 |
| | Number of STA nodes | [1, 60] |
| | Simulation time | 30 sec |
| | Event monitoring model | *FlowMonitor* |

|  | Animation model | *NetAnim* |
|---|---|---|
| Scenario 1 | RTS threshold | {0, 65535} |
| Scenario 2 | $CW_{min}$ | {3, 7, 15, 31} |
| Scenario 3 | $CW_{max}$ | {255, 511, 1023} |
| Scenario 4 | Retry limit | {1, 3, 5, 7, 9, 11} |

### 5.1. Effects of Access Mechanism

In the scenario 1, we consider the WLAN configured as Section 5 and we separately execute the experiment with basic access mechanism and RTS/CTS access mechanism which is controlled by RTS threshold value. The effect of access mechanisms (basic and RTS/CTS) on PDR, PLR, aggregated throughput and average delay under a varying number of STA nodes is illustrated in Figure 2. It is noted that PDL, PLR and average delay become stable and aggregated throughput increases linearly up to 12 nodes for both basic and RTS/CTS access mechanism. This is because the small number STA nodes lead to the non-saturation condition. With the number of STA nodes increasing within a certain range, it would not cause more collisions, which resulting in a stable PDL, PLR, and average delay and an increase of aggregated throughput. With the number of STA nodes further increasing, the network leads to saturation condition, more STA nodes will contend for transmission simultaneously, which would result in more collisions and thus an exponential decrease of PDR, a linear decrease of aggregated throughput, an exponential increase of PLR, and average delay. Moreover, it is also seen that the RTS/CTS access mechanism significantly improves the network performance compare to the basic access mechanism especially when the network is saturated. This is because the RTS/CTS access overcomes the hidden terminal problem and thus a less collision. In addition, the duration of collision transmission of RTS/CTS is less than basic access which offers improved network performance.

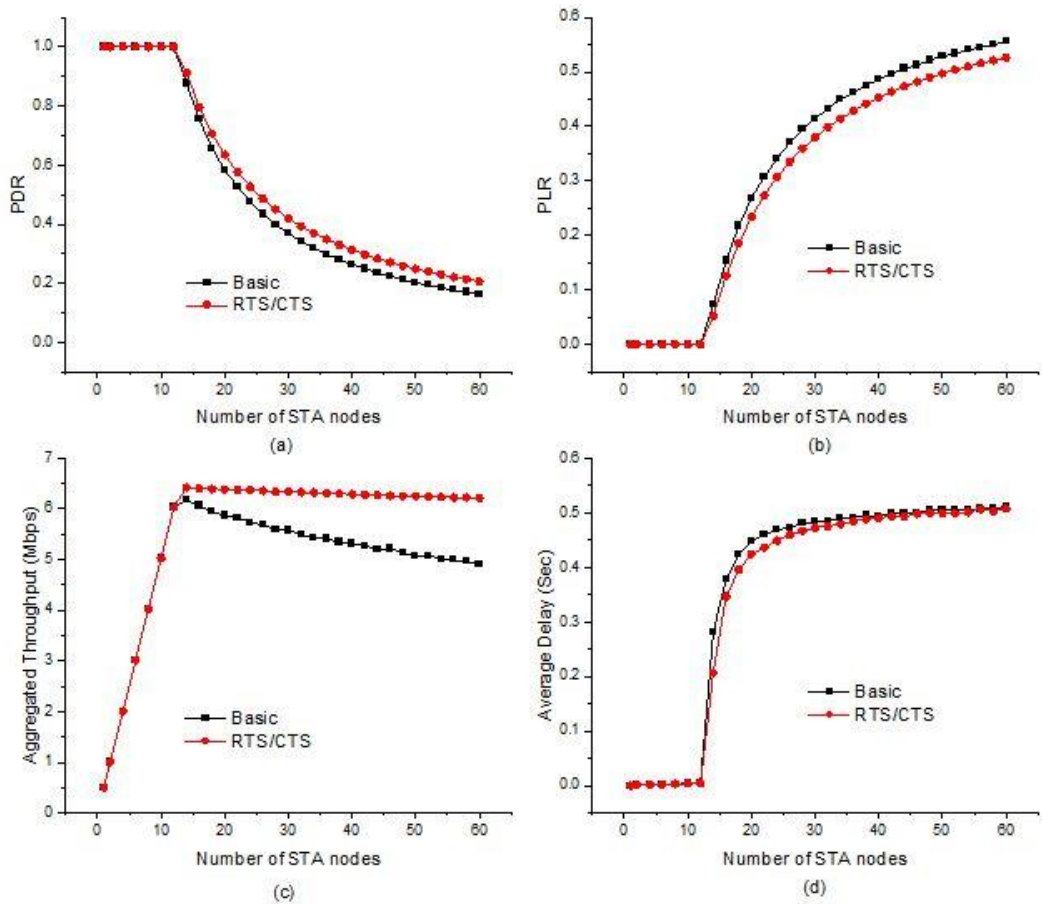

Figure 2: Effects of access mechanism on (a) packet delivery ratio (PDR), (b) packet lost ratio (PLR), (c) aggregated throughput, and (d) average delay

## 5.2. Effects of Minimum Contention Window ($CW_{min}$)

In the scenario 2, we consider the WLAN configured as Section 5 and we set the $CW_{min}$ value 3, 7, 15 and 31 using *Set ()* function of *Config* class. The effect of $CW_{min}$ on PDR, PLR, aggregated throughput and average delay under a varying number of STA nodes is illustrated in Figure 3. It is noted that PDL, PLR and average delay become stable and aggregated throughput increases linearly up to 10 nodes for all $CW_{min}$. For small number of STA nodes, the network is in non-saturated state. With the number of STA nodes increasing within a certain range, it would not cause more collisions, which resulting in a stable PDL, PLR, and average delay and an increase of aggregated throughput for all $CW_{min}$. Moreover, it is also seen that lower value of $CW_{min}$ leads to low PDR and aggregated throughput and high PLR and average delay when the network is in saturated state. This is because a lower value of $CW_{min}$ would result in more collisions than higher value of $CW_{min}$ and thus a decrease of PDR and aggregated throughput, an increase of PLR and average delay.

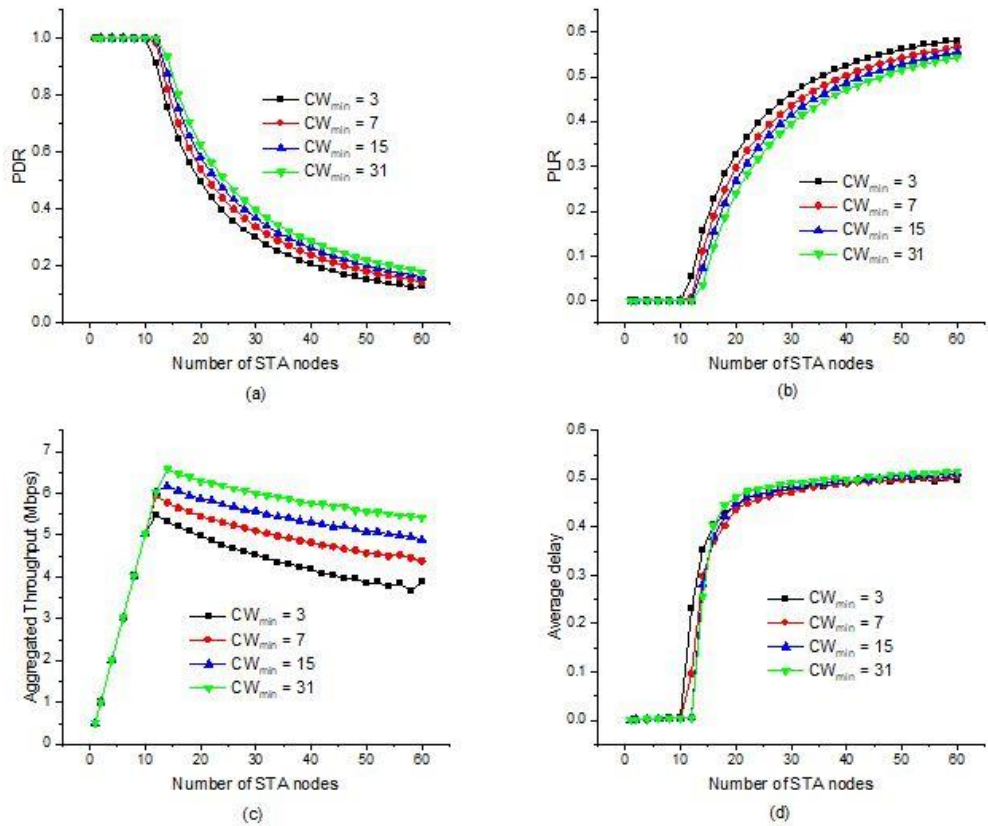

Figure 3: Effects of minimum contention window on (a) packet delivery ratio (PDR), (b) packet lost ratio (PLR), (c) aggregated throughput, and (d) average delay

## 5.3. Effects of Maximum Contention Window ($CW_{max}$)

In the scenario 3, we consider the WLAN configured as Section 5 and we set the $CW_{max}$ value 255, 511 and 1023 using *Set ()* function of *Config* class. The effect of $CW_{max}$ on PDR, PLR, aggregated throughput and average delay under a varying number of STA nodes is illustrated in Figure 4. It is also noted that PDL, PLR and average delay become stable and aggregated throughput increases linearly up to 10 nodes for all $CW_{max}$. Moreover, it is also seen that higher value of $CW_{max}$ leads to high PDR and aggregated throughput and low PLR and average delay when the network is in saturated state. This is because a higher value of $CW_{max}$ would result in less collision than lower value of $CW_{max}$ and thus an increase of PDR and aggregated throughput, a decrease of PLR and average delay. Unlike $CW_{min}$, $CW_{max}$ has no significant impact to change the value of performance metrics.

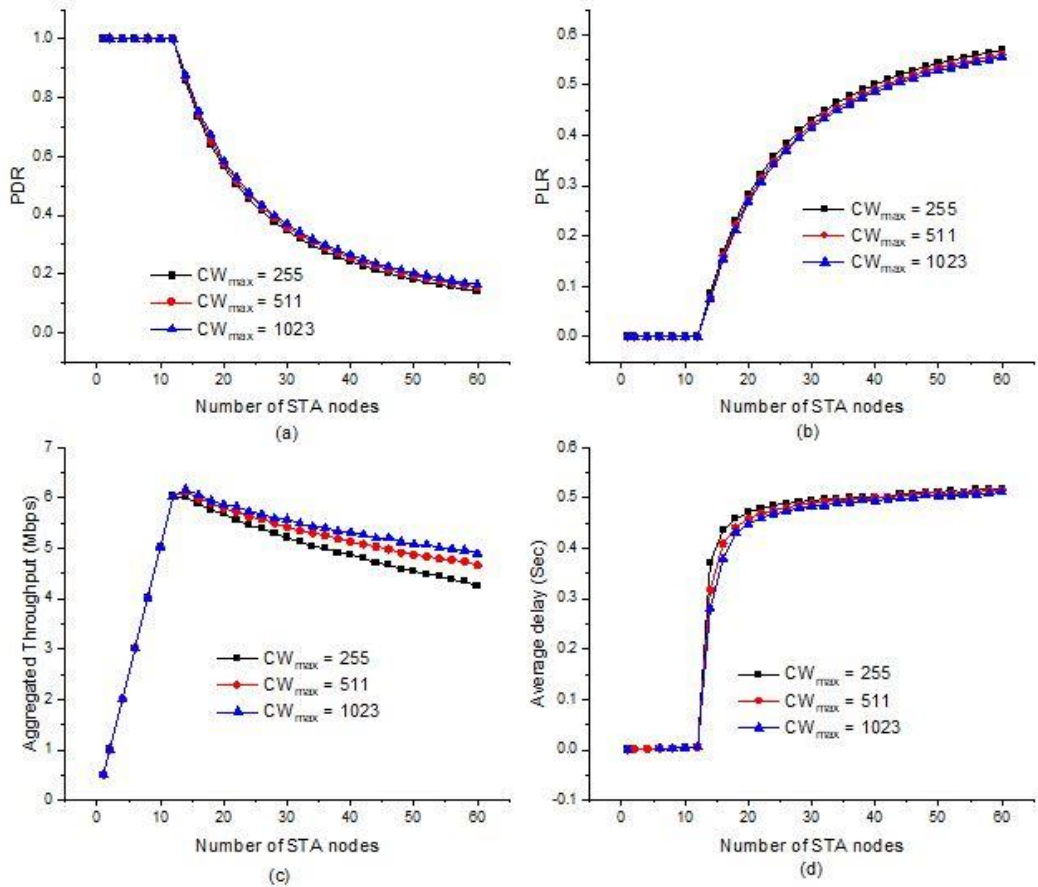

Figure 4: Effects of maximum contention window on (a) packet delivery ratio (PDR), (b) packet lost ratio (PLR), (c) aggregated throughput, and (d) average delay

### 5.4. Effects of Retry Limit

In the scenario 4, we consider the WLAN configured as Section 5 and we set the retry limit value 1, 3, 5, 7, 9 and 11 using Set () function of *Config* class. The effect of minimum contention window on PDR, PLR, aggregated throughput and average delay under a varying number of STA nodes is illustrated in Figure 5. It is noted that PDL, PLR, aggregated throughput and average delay become unstable for retry limit value 1 whereas the performance metrics significantly stable for retry value 3 to 11. Moreover, we also observed that there is no significant change of performance metrics for retry limit 7 to 11. This is because STA nodes no need to reach more than 7 to transmit packet successfully.

According to the results of this study, we can understand how different MAC parameters viz. access mechanisms, $CW_{min}$, $CW_{max}$, and retry limit, can affect the network performance metrics. The major findings of this study could be listed as follows:

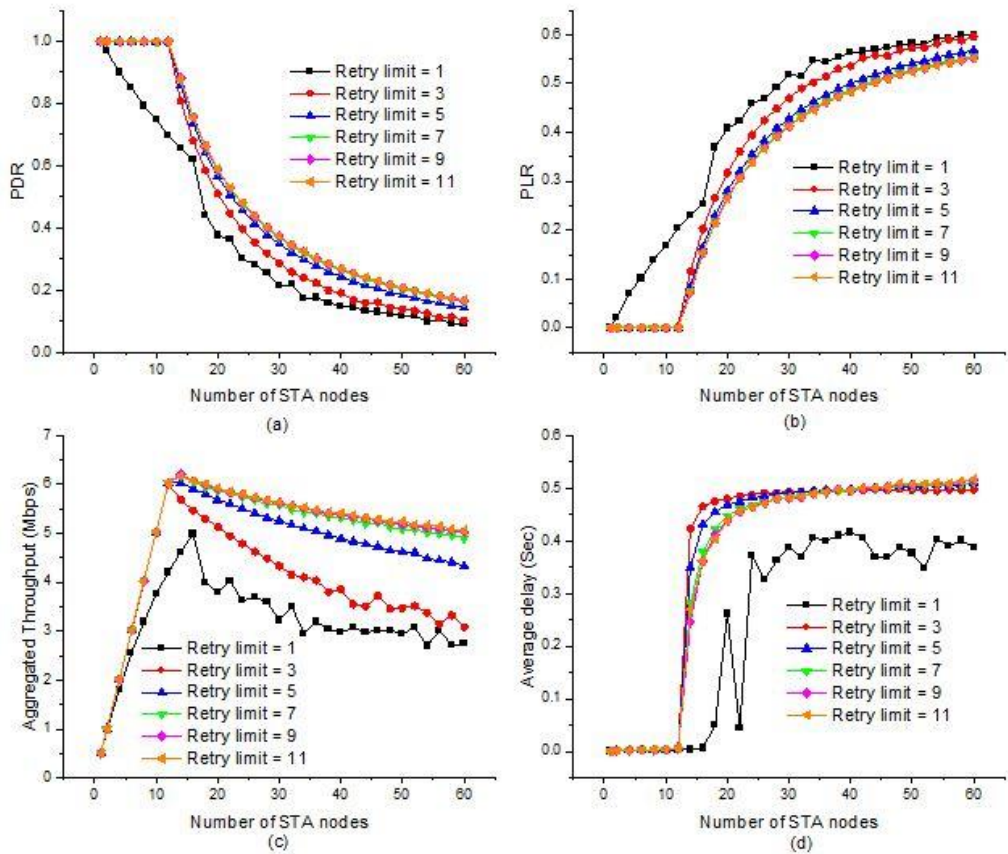

Figure 5: Effects of retry limit on (a) packet delivery ratio (PDR), (b) packet lost ratio (PLR), (c) aggregated throughput, and (d) average delay

i. When the network is in non-saturated state, there is no significant impact of MAC parameters on PDL, PLR, and average delay but the aggregated throughput linearly increases as the number of STA nodes increases.
ii. When the network is in saturated state, the performance of the network degrades as the number of STA nodes increases.
iii. The RTS/CTS access mechanism outperforms basic access mechanism, especially when the network is saturated.
iv. The higher value of minimum contention window lead to high throughput and average delay in saturated condition.
v. There is no remarkable impact of maximum contention window on the performance metrics.
vi. More than 7 retry limits have no significant impact on the performance metrics.

## 6. CONCLUSIONS

In this paper, we designed a NS-3 script to investigate the effects of MAC parameters on performance metrics for IEEE 802.11a standard based WLAN in NS-3 and organized it according to a TCP/IP model. The determination of performance metrics in context of NS-3 attributes is also discussed in this work. Unlike the previous related works, this hierarchical design procedure of WLAN in NS-3 provides a complete guideline for new user of NS-3. Moreover, we investigated the effects of MAC parameters (access mechanism, $CW_{min}$, $CW_{max}$, and retry limit) on network performance metrics viz. PDR, PLR, aggregated throughput, and average delay. The simulation results show that the RTS/CTS access mechanism offers better aggregated throughput than basic access mechanism in expense of average delay for dense network. Higher minimum contention window value provides better performance when network is in saturated state. Moreover, $CW_{max}$ and retry limit above 7 has no significant impact on the performance of the network. In addition, it also found that our designed NS-3 script based on IEEE 802.11a standard reacted accurately to the effects as desired in all modified dimensions. We believe that the simulation results would assist the protocol developers to design effective and efficient protocols and to select optimal value of different MAC parameters to enhance the performance. In future, we will explore the effects of different MAC and PHY parameters on both DCF and EDCA mechanisms under existing IEEE 802.11 standard family.

## CONFLICTS OF INTEREST

The authors declare no conflict of interest.

## ACKNOWLEDGEMENTS

This work was supported by the Institute of Research and Training (IRT), Hajee Mohammad Danesh Science and Technology University (HSTU), Dinajpur, Bangladesh.

## APPENDIX

```
#include "ns3/yans-wifi-helper.h"
#include "ns3/ssid.h"
#include "ns3/mobility-helper.h"
#include "ns3/internet-stack-helper.h"
#include "ns3/ipv4-address-helper.h"
#include "ns3/packet-sink-helper.h"
#include "ns3/on-off-helper.h"
#include "ns3/core-module.h"
#include "ns3/flow-monitor-module.h"
#include "ns3/netanim-module.h"

// This is a simple NS-3 script in order to show how to configure an IEEE 802.11a WLAN network and to
// investigate the effects of CWmin. It use CWmin value (3) and contains STA nodes varies from 1 to 60. It
// outputs the PDR, PLR, aggregated throughput, and average delay.

#define PI 3.14159265
using namespace ns3;
using namespace std;

NS_LOG_COMPONENT_DEFINE ("ieee802.11a");
int main ()
{
uint32_t apNumber = 1;
double simulationTime = 30; //seconds
bool useRtsCts = false;
uint16_t maxStaNumber = 60;
uint16_t minStaNumber = 1;
uint16_t nodeInterval = 2;
uint16_t padSta = 1 ;
uint16_t runNumber = maxStaNumber / nodeInterval;
uint16_t staNumber = minStaNumber;

string outputFilename = "effects of minimum CW.csv";
```

```
std::ofstream outputStream (outputFilename.c_str (), ios::app);
outputStream << "staNumber" << ",";
outputStream << "PDR" << ",";
outputStream << "PLR" << ",";
outputStream << "aggThroughput" << ",";
outputStream << "averageDelay" << endl;

for (uint32_t j = 0 ; j <= runNumber; j++)
         {
/*--------------Begin of PHY Topology Configuration ------------*/
NodeContainer apNode, staNodes;
apNode.Create (apNumber);
staNodes.Create (staNumber);
MobilityHelper apMobility, staMobility;
Ptr<ListPositionAllocator> positionAllocAp = CreateObject<ListPositionAllocator> ();
positionAllocAp->Add (Vector (0.0,0.0,0.0));
apMobility.SetPositionAllocator (positionAllocAp);
apMobility.SetMobilityModel ("ns3::ConstantPositionMobilityModel");
apMobility.Install (apNode);
Ptr<ListPositionAllocator> positionAlloc = CreateObject<ListPositionAllocator> ();
double distanceApToSta = 10;
for (uint32_t i = 0; i <= staNumber; i++)
         {
double theta = i * 2 * PI / staNumber;
positionAlloc->Add (Vector (distanceApToSta * cos(theta), distanceApToSta * sin(theta), 0.0));
         }
staMobility.SetPositionAllocator (positionAlloc);
staMobility.SetMobilityModel ("ns3::ConstantPositionMobilityModel");
staMobility.Install (staNodes);
/*-----------End of PHY Topology Configuration -------------*/

/*------Begin of Network Device Configuration -------*/
WifiHelper wifi;
wifi.SetStandard (WIFI_PHY_STANDARD_80211a);
wifi.SetRemoteStationManager ("ns3::ConstantRateWifiManager",
"DataMode", StringValue ("OfdmRate12Mbps"),
"ControlMode", StringValue ("OfdmRate12Mbps"),
"RtsCtsThreshold", UintegerValue (useRtsCts ? 0 : 65535) );

/*-------------Begin of PHY Layer Configuration ------------------*/
YansWifiChannelHelper wifiChannel = YansWifiChannelHelper::Default ();
wifiChannel.AddPropagationLoss ("ns3::RangePropagationLossModel");
YansWifiPhyHelper wifiPhy = YansWifiPhyHelper::Default ();
wifiPhy.SetErrorRateModel ("ns3::YansErrorRateModel");
wifiPhy.SetChannel (wifiChannel.Create ());
/* -------------------End of PHY Layer Configuration -------------*/

/* ----------------Begin of MAC Layer Configuration --------------*/
WifiMacHelper apWifiMac, staWifiMac;
Ssid ssid = Ssid ("ieee802.11a");
apWifiMac.SetType ("ns3::ApWifiMac", "Ssid", SsidValue (ssid));
staWifiMac.SetType ("ns3::StaWifiMac", "Ssid", SsidValue (ssid));
/* -------------End of MAC Layer Configuration -------------*/

NetDeviceContainer apNetDevice, staNetDevices;
apNetDevice = wifi.Install (wifiPhy, apWifiMac, apNode);
staNetDevices = wifi.Install (wifiPhy, staWifiMac, staNodes);
```

International Journal of Wireless & Mobile Networks (IJWMN), Vol.13, No.6, December 2021*/*------------End of Network Device Configuration ------------------*/*

*/*------------Begin of Modification of Attributes value ------------*/*
Config::Set ("/NodeList/*/$ns3::Node/DeviceList/*/$ns3::WifiNetDevice/Mac/$ns3::ApWifiMac/Txop/$ns3::Txop/MinCw", UintegerValue (3));
Config::Set ("/NodeList/*/$ns3::Node/DeviceList/*/$ns3::WifiNetDevice/Mac/$ns3::StaWifiMac/Txop/$ns3::Txop/MinCw", UintegerValue (3));
*/*------------End of Configuration of Attributes value --------*/*

*/*---------Begin of Network and Transport Layer Configuration -------*/*
InternetStackHelper internetStack;
internetStack.Install (apNode);
internetStack.Install (staNodes);
Ipv4AddressHelper ipv4Address;
ipv4Address.SetBase ("192.168.1.0", "255.255.255.0");
Ipv4InterfaceContainer staNodeInterface, apNodeInterface;
apNodeInterface = ipv4Address.Assign (apNetDevice);
staNodeInterface = ipv4Address.Assign (staNetDevices);
*/*---------End of Network and Transport Layer Configuration -------*/*

*/*---------Begin of Application Layer Configuration -------*/*
ApplicationContainer staApplications, apApplications;
uint32_t apPortNumber = 9;
for (uint32_t staIndex = 0; staIndex < staNumber; ++staIndex)
{
InetSocketAddress apSocketAddress (apNodeInterface.GetAddress (0), apPortNumber++);
OnOffHelper staOnOff ("ns3::UdpSocketFactory", apSocketAddress);
staApplications.Add (staOnOff.Install (staNodes.Get (staIndex)));
PacketSinkHelper apSink ("ns3::UdpSocketFactory", apSocketAddress);
apApplications.Add (apSink.Install (apNode.Get (0)));
}
apApplications.Start (Seconds (0.0));
apApplications.Stop (Seconds (simulationTime+1));
staApplications.Start (Seconds (1.0));
staApplications.Stop (Seconds (simulationTime+1));
*/*---------End of Application Layer Configuration -------*/*

*/*----------Begin of Animation Setup --------*/*
AnimationInterface animation ("ieee802.11a-animation.xml");
uint32_t uavIcon = animation.AddResource ("/home/abubakar/Desktop/ns-allinone-3.30.1/netanim-3.108/ap.png");
animation.UpdateNodeImage (0, uavIcon);
animation.UpdateNodeSize (0, 4, 4);
uint32_t nodeIcon = animation.AddResource ("/home/abubakar/Desktop/ns-allinone-3.30.1/netanim-3.108/sta.png");
for (uint16_t k = 1; k <= staNumber; k++)
    {
animation.UpdateNodeImage (k, nodeIcon);
animation.UpdateNodeSize (k, 2, 2);
    }
animation.SetMaxPktsPerTraceFile(1000000);
*/*----------End of Animation Setup --------*/*

*/*------ Begin of Event Monitoring --------*/*
NS_LOG_INFO ("Setup flow monitor.");

```
FlowMonitorHelper monitor;
Ptr<FlowMonitor> flowMonitor = monitor.InstallAll ();
flowMonitor->SerializeToXmlFile("ieee802.11a-flowmonitor.xml", true, true);
/*------ End of Event Monitoring --------*/

Simulator::Stop (Seconds (simulationTime + 1));
Simulator::Run ();
/*------ Begin of Data Collection --------*/
flowMonitor->CheckForLostPackets ();
Ptr<Ipv4FlowClassifier> classifier = DynamicCast<Ipv4FlowClassifier> (monitor.GetClassifier ());
std::map<FlowId, FlowMonitor::FlowStats> stats;
std::map<FlowId, FlowMonitor::FlowStats>::const_iterator i;
stats = flowMonitor->GetFlowStats ();
double totalTransmitedPackets = 0;
double totalReceivedPackets = 0;
double totalLostPackets = 0;
double aggThroughput = 0;
double totalDelay = 0;
for (i = stats.begin (); i != stats.end (); ++i)
{
double throughputPerFlow = i->second.rxBytes * 8.0 / simulationTime / 1000 / 1000;
aggThroughput += throughputPerFlow;
double transmitedPacketsPerFlow = i->second.txPackets ;
totalTransmitedPackets += transmitedPacketsPerFlow;
double receivedPacketsPerFlow = i->second.rxPackets ;
totalReceivedPackets += receivedPacketsPerFlow;
double lostPacketsPerFlow = i->second.lostPackets ;
totalLostPackets += lostPacketsPerFlow;
double delayPerFlow = i->second.delaySum.GetSeconds () ;
totalDelay += delayPerFlow;
}
double packetDeliveryRatio = totalReceivedPackets/ totalTransmitedPackets ;
double packetLostRatio = totalLostPackets / totalTransmitedPackets ;
double averageDelay = totalDelay / totalReceivedPackets ;
outputStream << staNumber << ",";
outputStream << packetDeliveryRatio << ",";
outputStream << packetLostRatio << ",";
outputStream << aggThroughput << ",";
outputStream << averageDelay << endl;
/*------ End of Data Collection --------*/
Simulator::Destroy ();
NS_LOG_INFO ("Destroyed.");
staNumber = staNumber + nodeInterval - padSta;
padSta = 0;
        }
outputStream.close ();
   return 0;
}
```

## AUTHORS

**Md. Abubakar Siddik** received the B. Sc. from the department of Telecommunication and Electronic Engineering (currently ECE), Hajee Mohammad Danesh Science and Technology University (HSTU), Dinajpur-5200, Bangladesh in 2011. He also received the M. Sc. from Institute of Information and Communication Technology (IICT), Bangladesh University of Engineering and Technology (BUET), Dhaka-1205, Bangladesh in 2017. He worked as a Lecturer in Prime University, Dhaka, Bangladesh and City University, Dhaka, Bangladesh from 2013-2014. He joined as a Lecturer in department of Electronics and Communication Engineering (ECE), HSTU in 2014 and currently he is working as an Assistant Professor in the same department. His current research interests include
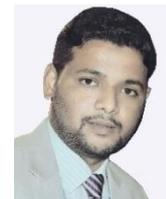

communication protocol of IoT, Smart Grid, UAV, UWSN, WBAN, VANET, in particular, design and performance analysis of communication protocol via simulation and analytical model.

**Jakia Akter Nitu** received the B. Sc. (Engineering) from the department of Electronics and Communication Engineering (ECE), Hajee Mohammad Danesh Science and Technology University (HSTU), Dinajpur-5200, Bangladesh in 2019. Currently, she is a student of M. Sc. (Engineering) in the same department. Her research interest is performance analysis of communication protocols of IoT, UAV, WBAN and WSN. 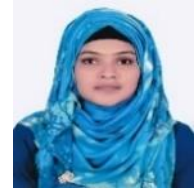

**Natasha Islam** received the B. Sc. (Engineering) from the Department of Electronics and Communication Engineering (ECE), Hajee Mohammad Danesh Science and Technology University (HSTU), Dinajpur-5200, Bangladesh in 2019. Currently, she is working as Product Operations Engineer at SquareFeet (A PropTech platform in Bangladesh). Her research interest is access mechanism of IoT and WSN. 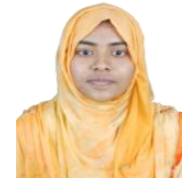

**Most. Anju Ara Hasi** received the B. Sc. (Engineering) from the department of Electronics and Communication Engineering (ECE), Hajee Mohammad Danesh Science and Technology University (HSTU), Dinajpur-5200, Bangladesh in 2017. Currently, she is a M. Sc. (Engineering) thesis semester student in the same department. Her research interest is performance analysis of communication protocols of IoT and WBAN. 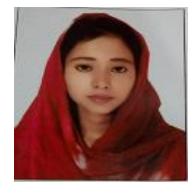

**Jannatun Ferdous** received the B. Sc. from the department of Telecommunication Electronic Engineering (currently ECE) Hajee Mohammad Danesh Science and Technology University (HSTU) Dinajpur-5200, Bangladesh in 2012. She joined as a Lecturer in the Dept. of Electronics and Communication Engineering (ECE), HSTU in 2016 and currently she is working as an Assistant Professor in the same department. Her current research interests include communication protocol IoT, Network Security. 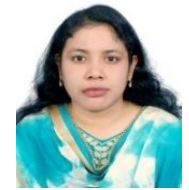

**Md. Mizanur Rahman** received the B. Sc. degree in Electrical and Electronic Engineering from the Rajshahi University of Engineering & Technology,Rajshahi-6204, Bangladesh in 2012.He worked as a Lecturer in the Department of Electrical and Electronic Engineering at Hajee Mohammad Danesh Science and Technology University,Dinajpur-5200,Bangladesh from 2013 to 2016.He has been serving as an Assistant Professor in the same university since 2016 and his research interests include Renewable Energy, Smart Grid System and Power System Stability. He is also the member of IEEE. 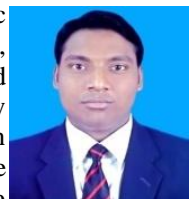

**Nahid Sultan** received the Bachelor of Science in Computer Science and Engineering from Islamic University (IU), kushtia-7003, Bangladesh in the year of 2011. He received M.Sc. in Computer Science and Engineering from Islamic University (IU), kushtia-7003, Bangladesh in the year of 2012. From February 2015 to September 2015 he worked on a Private University name UITS as a Lecturer in the country. In September 2015, he joined as a Lecturer in the department of Computer Science and Engineering, HSTU, Dinajpur-5200, Bangladesh. Currently he is working as an Assistant Professor in the same department. His research interest includes Artificial Intelligence, Machine Learning, Cognitive Radio Network, Image Processing, Bio-informatics, and Internet of Things. 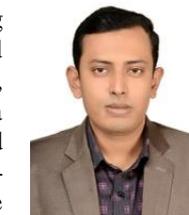